\documentclass[aps,twocolumn, pra, footinbib,superscriptaddress]{revtex4-1}
\usepackage{amsmath,amssymb} \usepackage{colordvi} \usepackage{color}
\usepackage{graphicx}
\usepackage{dcolumn}
\usepackage{bm}
\usepackage{romannum}
\usepackage{caption}
\usepackage{subcaption}
\usepackage{cleveref}
\usepackage{float}
\usepackage{csquotes}
\usepackage{braket}
\usepackage{verbatim}
\begin{document}

 \draft
 \title{Rabi spectroscopy of three-dimensional optical lattice clocks}

 \author{Guangcun Liu}
	\affiliation{Laboratory of Quantum Engineering and Quantum Metrology, School of Physics and Astronomy, Sun Yat-Sen University (Zhuhai Campus), Zhuhai 519082, China}
	\author{Yinan Huang}
	\affiliation{Laboratory of Quantum Engineering and Quantum Metrology, School of Physics and Astronomy, Sun Yat-Sen University (Zhuhai Campus), Zhuhai 519082, China}
	\author{Zhuo Cheng}
	\affiliation{Laboratory of Quantum Engineering and Quantum Metrology, School of Physics and Astronomy, Sun Yat-Sen University (Zhuhai Campus), Zhuhai 519082, China}
	\author{Ruize Chen}
	\affiliation{Laboratory of Quantum Engineering and Quantum Metrology, School of Physics and Astronomy, Sun Yat-Sen University (Zhuhai Campus), Zhuhai 519082, China}
	
	\author{Zhenhua Yu}
	\email[]{huazhenyu2000@gmail.com}
	\affiliation{Laboratory of Quantum Engineering and Quantum Metrology, School of Physics and Astronomy, Sun Yat-Sen University (Zhuhai Campus), Zhuhai 519082, China}
	\affiliation{State Key Laboratory of Optoelectronic Materials and Technologies,
Sun Yat-Sen University (Guangzhou Campus), Guangzhou 510275, China}

\date{\today}

\begin{abstract}
Recent realisation of three-dimensional optical lattice clocks circumvents short range collisional clock shifts which have been the bottle neck towards higher precision; the long range electronic dipole-dipole interaction between the atoms becomes the primary source of clock shift due to interatomic interactions. We study the Rabi spectroscopy of three-dimensional optical lattice clocks with unity filling. From the Lindblad equation governing the time evolution of the density matrix of the atoms, we derive the Bloch equations in the presence of the external Rabi driving laser field, and solve the equations approximately to the first order of the coupling strength of the dipole-dipole interaction between the atoms. We find that the clock shift equals to the product of the coupling strength, a factor determined by the parameters of the Rabi pulse, and another factor depending on the configuration of the three-dimensional optical lattice. Our result on the clock shift within the Rabi spectroscopy can be checked by measurement in future experiment.
\end{abstract}

\maketitle

\section{introduction}

Atomic clocks have been an active research subject for a long time because of their essential role in both research exploration and practical application. Obstacles to achieving higher precision have been overcome step by step through the expenditure to achieve a better atomic clock \cite{review1,reviewjun}. Recently, new efforts have been directed to fermionic atoms since the Pauli exclusion principle is expected to help to reduce the collisional clock shifts for fermionic atoms; 
one dimensional optical lattice of magic wavelength has also been used to suppress the Doppler effect, and collisional clock shifts have been diagnosed carefully \cite{proceedings,magicwave,1djapan,naturereview,campbell,gibble, rey, yu,latticeshift, comparison,instability,1d}. 

On the other hand, new development in ultracold atoms on the Mott-insulator transition in Bose-Einstein condensates and ultracold Fermi gases also inspired advances in atomic clocks \cite{bec,mott,insulating}. In a recent experiment \cite{3dlatticejun}, a degenerate Fermi gas of $^{87}$Sr atoms is loaded into a three-dimensional optical lattice of magic wavelength in the Mott-insulating regime. In this strongly correlated regime, atoms prefer to occupying different lattice sites, which mitigates the collisional clock shifts due to on-site interactions between atoms. This realization of fermionic atomic clocks in three-dimensional optical lattices of magic wavelength has pushed the measurement precision of atomic clocks to a new level. These development in atomic clocks also opens a new door for researches such as the search for variation of fundamental constants and atomic parity violation \cite{search}. 

However, atomic clocks in three-dimensional optical lattices still face shifts due to the long range electronic dipole-dipole interaction between atoms. Atomic clocks usually make use of two internal electronic states of each atom, which are called clock states. One of them is the ground state $|g\rangle$ and the other is the excited state $|e\rangle$. The energy difference between the clock states provides a standard for frequency, and therefore time. The off-diagonal matrix elements of the electronic dipole moment $\vec d$ of each atom in the subspace spanned by the two clock states are nonzero, i.e., $\langle e|\vec d|g\rangle\neq0$, such that external laser fields can be used to couple to the electronic dipole moment of atoms to generate a superposition of the two clock states for each atom, and subsequently the energy difference can be extracted out via the time evolution of the atomic clocks. The same nonzero off-diagonal matrix elements of the electronic dipole moment $\langle e|\vec d|g\rangle$ give rise to the long range dipole-dipole interaction between atoms. The master equation have been adopted to examine the effects of the long range dipole-dipole interaction on the Ramsey spectroscopy and spin-squeezing \cite{ramsey,spinsqueez}. 

The Rabi spectroscopy is another widely used experimental method to probe clock shifts \cite{campbell, yu}. 
In this work, we study the Rabi spectroscopy for a cloud of fermionic atoms with unity filling in a three-dimensional optical lattice. We start with the Lindblad equation governing the time evolution of the density matrix of the atoms, and derive the Bloch equations in the presence of the external Rabi driving laser field. We solve the Bloch equations to the first order of the coupling strength of the dipole-dipole interaction between atoms. We find that the clock shift equals to the product of the coupling strength, a factor determined by the parameters of the Rabi pulse, and another factor depending on the configuration of the three-dimensional optical lattice. Being specific, for $^{87}$Sr, the resultant clock shift yields typically a relative precision of $10^{-18}$. By tuning the configuration of the three-dimensional optical lattice, one can further mitigate the clock shift due to the dipole-dipole interaction thanks to the Brag scattering. Our result on the clock shift in the Rabi spectroscopy can be compared with measurement in future experiment.

\section{The Lindblad Equation} 
We consider the situation that there is at most a single atom on each lattice site in the three-dimensional optical lattice clock, such that short-range collisional shifts become absent, and the primary interatomic interaction is due to the dipole-dipole interaction \cite{3dlatticejun}. During the time evolution of the clock, the electronic state of each atom is generally a superposition of the two clock states, the ground state $|g\rangle$ and the excited state $|e\rangle$. The off-diagonal matrix elements of the electronic dipole moment $\vec d$ of each atom in the subspace spanned by the two clock states are nonzero, i.e., $\langle e|\vec d|g\rangle\neq0$, which gives rise to the dipole-dipole interaction between the atoms. In addition, the same nonzero off-diagonal matrix elements of the electronic dipole moment is also the source for the finite (though usually quite long) lifetime of the excited clock state. Thus the interaction and the lifetime shall be treated on the same footing \cite{ramsey}. 

The Lindblad equation is an ideal formalism to take into account dissipation \cite{api}. Starting from the atoms coupled to a radiation field via their electronic dipole moments and assuming the radiation field relaxes rapidly to its vacuum state, previous works have derived the Lindblad equation applicable to our problem as
 \cite{essay,ramsey,protectedstate,spinsqueez}
\begin{equation}
\frac{\partial\rho}{\partial t}=\frac{1}{i\hbar}[H_0+H_{dd}+H_{drive},\rho]+\mathcal{L}[\rho],\label{lindblad}
\end{equation}
where $\rho$ is the density matrix for the atoms. The free Hamiltonian is
\begin{equation}
H_0=\frac{\hbar\omega_0}{2}\sum_a\sigma_a^z
\end{equation}
and the dipole-dipole interaction Hamiltonian is
\begin{equation}
H_{dd}=\frac{\hbar\Gamma}{2}\sum_{a,b(a\neq b)}g(k'\vec{r}_{ab})\sigma_a^+\sigma_b^-.\label{dd}
\end{equation}
Coupling between the atoms and external driving fields is described by $H_{drive}$. 
The dissipation is captured by
\begin{equation}
\mathcal{L}[\rho]=-\frac{\Gamma}{2}\sum_{a,b}f(k'\vec{r}_{ab})(\lbrace\sigma_a^+\sigma_b^-,\rho\rbrace-2\sigma_b^-\rho\sigma_a^+).\label{l}
\end{equation}
For the $a$th atom, the pesudo-spin operators $\vec \sigma_a$ are defined as $\sigma_a^z=|e_a\rangle\langle e_a|-|g_a\rangle\langle g_a|$, $\sigma_a^+=|e_a\rangle\langle g_a|$ and $\sigma_a^-=|g_a\rangle\langle e_a|$. The wavelength $k'$ is resonant with respect to the transition $|g\rangle\leftrightarrow|e\rangle$, i.e., $ck'=\hbar\omega_0$. The spontaneous decay rate of the excited clock state is $\Gamma$. The magnitude of $H_{dd}$ is also proportional to $\Gamma$ because the dipole-dipole interaction is due to the exchange of the radiation field between the atoms.

The explicit expressions of $f(\vec{v})$ and $g(\vec{v})$ are given by
\begin{equation}
\begin{aligned}
&f(\vec{v})=\frac{3}{2}\left[\sin^2\theta\frac{\sin v}{v}+(3\cos^2\theta-1)\left(\frac{\sin v}{v^3}-\frac{\cos v}{v^2}\right)\right] \\
&g(\vec{v})=-\frac{3}{2}\left[\sin^2\theta\frac{\cos v}{v}+(3\cos^2\theta-1)\left(\frac{\cos v}{v^3}+\frac{\sin v}{v^2}\right)\right],
\end{aligned}
\end{equation}
where $\theta$ is the angle between $\vec{v}$ and the dipole polarization direction. In case that the atomic clock is driven by an external linear polarized laser field, the dipole polarization direction should be parallel to that of the laser field. In the following calculation, we assume the dipole polarization is in the $z$ direction. 

In the long wavelength limit $k\to0$, the leading term of $H_{dd}$ is 
$(3\hbar\Gamma/4k'^3)\sum_{a\neq b}(1-3\cos^2\theta)\sigma_a^+\sigma_b^-/r_{ab}^3$, which is the familiar static form of the dipole-dipole interaction. In the same limit, the dissipation part becomes $\mathcal{L}[\rho]=-(N\Gamma/2)(\lbrace\Sigma^+\Sigma^-,\rho\rbrace-2\Sigma^-\rho\Sigma^+)$ with the total pseudo-spin $\vec\Sigma=\sum_{a=1}^N \vec\sigma_a/\sqrt N$; the long wavelength radiation field induces loss by coupling to the total pseudo-spin.

\section{the Rabi spectroscopy}\label{rabi}
The Rabi spectroscopy is a widely used method to extract shifts in atomic clocks \cite{campbell, yu}. The procedure of the spectroscopy is as following. 
The atoms are initially prepared in one of the two clock states, e.g., the ground state $\ket g$. A $\pi$-pulse of a linear polarized laser with frequency $\omega_L$ ($\approx\omega_0$) is applied to the atoms. After the pulse the fraction of the atoms transited to the other clock state $P$ is measured; Overall, $P$ shall be smaller for larger detuning $\delta\equiv\omega_L-\omega_0$. In the absence of interatomic interactions, $P$ shall be symmetric with respect to $\delta$, i.e., $P(\delta)=P(-\delta)$ \cite{campbell, yu}. Generally speaking, this symmetry is no longer respected when interatomic interactions set in. Thus experimentally, the clock shift $\delta_c$ can be determined by measuring one red detuning $\delta_R$ ($<0$) and one blue detuning $\delta_B$ ($>0$), which give rise to the same final transited atom fraction $P$, i.e., $P(\delta_R)=P(\delta_B)$; the clock shift is defined as $\delta_c=\delta_R+\delta_B$. For the sake of sensitivity, in experiment $\delta_R$ is usually chosen such that $P(\delta_R)$ has a large derivative.

The experimental observable $P(\delta)$ can be obtained theoretically by calculating the expectation value of the $z$ component of the total pseudospin
\begin{equation}
\langle\Sigma_z\rangle={\rm Tr}\left(\sum_a\sigma_a^z\rho(t)\right)=\sum_a\left\langle\sigma_a^z\right\rangle. \\
\end{equation}
If one starts with all the atoms in $|g\rangle$, at the end of the pulse, the transited fraction to the state $|e\rangle$ is $P=(1+\langle\Sigma_z\rangle/N)/2$. The time evolution of the density matrix is governed by the Lindblad equation (\ref{lindblad}), in which, for the Rabi spectroscopy, in the rotating frame after applying the rotating wave approximation, the drive Hamiltonian is
\begin{equation}
H'_{drive}=\sum_ai\hbar\Omega_0(\sigma_a^+e^{i\vec{k}\cdot\vec{r}_a}-\sigma_a^-e^{-i\vec{k}\cdot\vec{r}_a})\label{hd}
\end{equation}
and the free Hamiltonian becomes 
\begin{equation}
H'_0=-\frac{\hbar\delta}2\sum_a\sigma^z_a,\label{h0}
\end{equation} 
while $H'_{dd}=H_{dd}$ and $\mathcal{L}[\rho]$ is invariant.
Here $\vec{k}$ is the wave vector of the Rabi driving laser, and $\Omega_0$ is the Rabi frequency proportional to the nonzero off-diagonal matrix element of the electronic dipole moment $\langle e|d_z|g\rangle$, since we assume the polarization of the Rabi driving laser is along the $z$ direction. The combined Hamiltonian has the form
\begin{align}
H'_0+H'_{drive}=-\sum_a \vec B_a\cdot \vec\sigma_a,
\end{align}
with the effective field $\vec B_a=(\hbar \Omega_0\sin(\vec{k}\cdot\vec{r}_a),\hbar \Omega_0\cos(\vec{k}\cdot\vec{r}_a),\hbar\delta/2)$ in the Cartesian coordinates.
In the following calculation, we assume $\vec k$ is along the $x$ direction. Note that since we are interested in small detuning, i.e., $|\delta|\ll\omega_0$, to the lowest order we neglect the magnitude difference between $k$ and $k'$. 
For simplicity, we have also neglected the possible variation of $\Omega_0$ for different atoms due to the small spatial imhomogeneity in the driving laser intensity \cite{campbell, yu}.


\section{the Bloch equations}
We derive the Bloch equation governing the evolution of the pseudospin of each atom from the Lindblad formalism, Eq.~(\ref{lindblad}), and have
\begin{align}
\frac{\partial \left\langle\sigma_a^z\right\rangle}{\partial t}=&2\Omega_0[\left\langle\sigma_a^+\right\rangle e^{ikx_a}+\left\langle\sigma_a^-\right\rangle e^{-ikx_a}]-\Gamma\left\langle\sigma_a^z\right\rangle-\Gamma\nonumber\\
&-\Gamma\sum_{b (\neq a)}[f(k\vec{r}_{ab})+ig(k\vec{r}_{ab})]\left\langle\sigma_a^+\sigma_b^-\right\rangle\nonumber\\
&-\Gamma\sum_{b (\neq a)}[f(k\vec{r}_{ab})-ig(k\vec{r}_{ab})]\left\langle\sigma_a^-\sigma_b^+\right\rangle
,\label{z}
\end{align}
and
\begin{equation}
\begin{aligned}
\frac{\partial \left\langle\sigma_a^+\right\rangle}{\partial t}=&-i\delta\left\langle\sigma_a^+\right\rangle-\Omega_0\left\langle\sigma_a^z\right\rangle e^{-ikx_a} \\
&-\frac{\Gamma}{2}\left\langle\sigma_a^+\right\rangle+\frac{\Gamma}{2}\sum_{b(\neq a)}[f(k\vec{r}_{ab})-ig(k\vec{r}_{ab})]\left\langle\sigma_a^z\sigma_b^+\right\rangle.
\end{aligned}\label{plus}
\end{equation}
The equation of motion for $\left\langle\sigma_a^-\right\rangle$ can be obtained by taking the complex conjugate of Eq.~(\ref{plus}).

To solve for $\langle\sigma_a^i\rangle$, one needs to determine the behavior of second order correlation functions $\langle\sigma_a^i \sigma_b^j\rangle$ for $b\neq a$, whose equation of motion likewise depends on third order correlation functions $\langle\sigma_a^i \sigma_b^j\sigma_c^\ell\rangle$. This dependence of lower order correlation functions on higher order ones continue to order $N$ where $N$ is the total number of atoms. 
Since the excited clock state is usually very long lived, i.e., $\Gamma$ is small, we solve Eqs.~(\ref{z}) and (\ref{plus}) perturbatively to the first order of $\Gamma$. We define $\tilde\sigma_a^+=\sigma_a^+ e^{ikx_a}$, $\tilde\sigma_a^z=\sigma_a^z$, $\tilde\sigma_a^x=\tilde\sigma_a^++\tilde\sigma_a^-$ and $\tilde\sigma_a^y=-i(\tilde\sigma_a^+-\tilde\sigma_a^-)$; at $t=0$, the initial wavefunction is $\prod_{a=1}^N|g_a\rangle$ and
\begin{align}
\left\langle\tilde\sigma_a^z(0)\right\rangle=&-1,\\
\left\langle \tilde\sigma_a^+(0)\right\rangle=&0.\label{initial}
\end{align}
Thus we cast Eqs.~(\ref{z}) and (\ref{plus}) into the following form
\begin{align}
\left(\frac{\partial}{\partial t}-\mathbf Q\right)\begin{bmatrix} \langle\tilde\sigma_a^x\rangle\\
\langle\tilde\sigma_a^y\rangle\\
\langle\tilde\sigma_a^z\rangle\end{bmatrix}
=\Gamma \begin{bmatrix} F_a^++F_a^-\\
-i(F_a^+-F_a^-)\\
F_a^z\end{bmatrix},\label{m}
\end{align}
where
\begin{align}
\mathbf Q=2\Omega \begin{bmatrix} 0 & \cos\phi & -\sin\phi\\
-\cos\phi & 0 & 0\\
\sin\phi & 0 & 0
\end{bmatrix},
\end{align}
and 
\begin{align}
F^+_a=&-\frac12\langle\tilde\sigma_a^+\rangle_0+\frac12\sum_{b(\neq a)}h_{ab}\langle\tilde\sigma_a^z\rangle_0\langle\tilde\sigma_b^+\rangle_0\\
F^z_a=&-1-\langle\tilde\sigma_a^z\rangle_0-\sum_{b(\neq a)}h_{ab}\langle\tilde\sigma_a^-\rangle_0\langle\tilde\sigma_b^+\rangle_0\nonumber\\
&-\sum_{b(\neq a)}h^*_{ab}\langle\tilde\sigma_a^+\rangle_0\langle\tilde\sigma_b^-\rangle_0\\
F^-_a=&(F^+_a)^*
\end{align}
with
\begin{align}
h_{ab}=&[f(k\vec{r}_{ab})-ig(k\vec{r}_{ab})]e^{ik(x_a-x_b)},\\
\Omega=&\sqrt{\Omega_0^2+\delta^2/4},
\end{align}
and $\cos\phi=\delta/2\Omega$ and $\sin\phi=\Omega_0/\Omega$.

The zero order solutions $\langle\vec\sigma_a\rangle_0$ are obtained by taking $\Gamma=0$ in Eq.~(\ref{m}), and have the explicit forms
\begin{equation}
\begin{aligned}
\left\langle\tilde\sigma_a^+(t)\right\rangle_0 =&\frac12\sin\phi\sin(2\Omega t)-\frac i2\sin\phi\cos\phi[1-\cos(2\Omega t)],\\
\left\langle\tilde\sigma_a^z(t)\right\rangle_0=&-\cos^2\phi-\sin^2\phi\cos(2\Omega t).
\end{aligned}
\end{equation}
When neglecting the electric dipole coupling between the two clock states, after a $\pi$-pulse, i.e., $2\Omega_0 t=\pi$, the transited fraction of atoms to the excited state is $P(\delta)=\sin^2\left((\pi/2)\sqrt{1+(\delta/2\Omega_0)^2}\right)/[1+(\delta/2\Omega_0)^2]$, which is symmetric in $\delta$ as shown in Fig. \ref{noninteractrabi}.

To solve Eq.~(\ref{m}) to first order of $\Gamma$, we employ the Green's function $\mathbf G(t)$ which satisfies \cite{yu}
\begin{align}
\left(\frac{\partial}{\partial t}-\mathbf Q\right)\mathbf G(t)=\delta(t)
\end{align}
with the initial condition $\mathbf G(0^+)=1$ and $\mathbf G(0^-)=0$. In the matrix form
\begin{align}
\mathbf G(t)=\theta(t)\mathbf R\begin{bmatrix} 1 &0 &0\\
0& e^{-i2\Omega t} &0\\
0&0&e^{i2\Omega t}\end{bmatrix}\mathbf R^\dagger,
\end{align}
with
\begin{align}
\mathbf R=\frac1{\sqrt2}\begin{bmatrix} 0 &i &-i\\
\sqrt2\sin\phi& \cos\phi &\cos\phi\\
\sqrt2\cos\phi&-\sin\phi &-\sin\phi \end{bmatrix},
\end{align}
Thus from Eq.~(\ref{m}), we have
\begin{align}
\begin{bmatrix} \langle\tilde\sigma_a^x(t)\rangle\\
\langle\tilde\sigma_a^y(t)\rangle\\
\langle\tilde\sigma_a^z(t)\rangle\end{bmatrix}
=&\mathbf G(t)\begin{bmatrix} \langle\tilde\sigma_a^x(0)\rangle\\
\langle\tilde\sigma_a^y(0)\rangle\\
\langle\tilde\sigma_a^z(0)\rangle\end{bmatrix}\nonumber\\
&+\Gamma\int^t_0 dt' \mathbf G(t-t')
 \begin{bmatrix} F_a^+(t')+F_a^-(t')\\
-i[F_a^+(t')-F_a^-(t')]\\
F_a^z(t')\end{bmatrix}.
\end{align}
We find explicitly
\begin{widetext}
\begin{align}
\langle\tilde\sigma_a^z\rangle=&\left\langle\tilde\sigma_a^z\right\rangle_0+\frac{\Gamma\sin^2(\phi)}{2\Omega}
\left\{-\frac{1}{8} \left[6 \sin (\tau ) \sin ^2(\phi )+4 \tau  \cos ^2(\phi )+\tau  \cos (\tau ) (\cos (2 \phi )-5)\right]+{\rm Im}\left(\sum_{b(\neq a)}h_{ab}\right)\right.\nonumber\\
& \times\left[\frac{1}{4} \cos (\phi ) \left[\sin (\tau ) (\sin (\tau )-\tau ) \sin ^2(\phi )+2 \cos ^2(\phi ) (\tau  \sin (\tau )+2 \cos (\tau )-2)\right]\right]\nonumber\\
&\left.-{\rm Re}\left(\sum_{b(\neq a)}h_{ab}\right)\sin ^2\left(\frac{\tau }{2}\right)\left[\sin(\tau ) \sin ^2(\phi )+\tau  \cos ^2(\phi )\right]\right\}\label{first}
\end{align}
with $\tau=2\Omega t$,  and ${\rm Im}\left(\sum_{b(\neq a)}h_{ab}\right)=f(k\vec{r}_{ab})\sin(kx_{ab})-g(k\vec{r}_{ab})\cos(kx_{ab})$,  and ${\rm Re}\left(\sum_{b(\neq a)}h_{ab}\right)=f(k\vec{r}_{ab})\cos(kx_{ab})+g(k\vec{r}_{ab})\sin(kx_{ab})$.
\end{widetext}

\section{Clock Shift} 

\begin{figure}[H]
\centering
\includegraphics[scale=0.5]{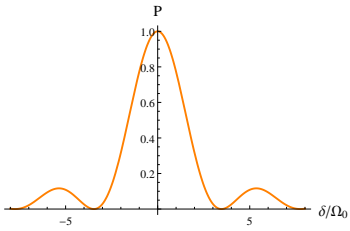}
\caption{The excitation fraction $P$ versus the detuning $\delta$ for free atoms after a $\pi$-pulse, i.e.,  $2\Omega_0 t=\pi$. }
\label{noninteractrabi}
\end{figure}

Compared with the non-interacting case in which the excitation fraction $P(\delta)$ after a $\pi$-pulse is symmetric in $\delta$ as shown in Fig.\ref{noninteractrabi}, the dipole-dipole interaction gives rise to terms of order $\Gamma$ in the expression of $\left\langle\tilde\sigma_a^z\right\rangle$ and makes the profile of $P(\delta)$ asymmetric; the maximum value of $P(\delta)$ thus shall shift to a nonzero value of $\delta$. 

We quantify the value of the clock shift $\delta_c$ in the way mentioned in Sec. \ref{rabi} as 
\begin{align}
\langle\Sigma_z(-\delta)\rangle=\langle\Sigma_z(\delta+\delta_c)\rangle,
\end{align}
where $\delta$ is a reference frequency which shall be chosen to enhance the sensitivity $\partial \langle\Sigma_z\rangle/\partial\delta$.
Since usually $\delta/\Omega_0$ chosen to be $\sim1$ (c.f. Fig. \ref{noninteractrabi}) and $\delta_c\ll\delta$, to the lowest order, we have \cite{yu}
\begin{align}
\delta_c=-\frac{\langle\Sigma_z(\delta)\rangle-\langle\Sigma_z(-\delta)\rangle}{\partial \langle\Sigma_z\rangle/\partial\delta}\label{dc}
\end{align}
From Eq. (\ref{first}), 
\begin{align}
&\langle\Sigma_z(\delta)\rangle-\langle\Sigma_z(-\delta)\rangle\nonumber\\
=&\frac{\Gamma}{4\Omega} {\rm Im}\left(\sum_{a,b(a\neq b)}h_{ab}\right)\sin^2(\phi) \cos (\phi ) \left[\sin (\tau ) (\sin (\tau )-\tau )  \right.\nonumber\\
&\left.\times\sin ^2(\phi )
+2 \cos ^2(\phi ) (\tau  \sin (\tau )+2 \cos (\tau )-2)\right].
\end{align}
On the other hand, we approximate $\partial \Sigma_z/\partial\delta$ to zero order of $\Gamma$ and find

\begin{align}
\frac{\partial \langle\Sigma_z\rangle}{\partial\delta}=\frac N{2\Omega} \sin ^2(\phi ) \cos (\phi ) [\tau  \sin (\tau )+2 \cos (\tau )-2].
\end{align}
Thus after a duration $\tau(=2\Omega t)$, 
\begin{align}
\delta_c/\Gamma=& A(\phi,\tau)F_{dd}\label{dc}
\end{align}
with
\begin{align}
A(\phi,\tau)=&\sin ^2(\phi ) \frac{[\tau-\sin (\tau)] \sin (\tau)}{2[ \cos (\tau)+ \tau \sin (\tau)-2]}-\cos ^2(\phi )\label{axy}
\end{align}
and 
\begin{equation}
F_{dd}=\frac{1}{N}\sum_{a,b(a\neq b)}[g(k\vec{r}_{ab})\cos(kx_{ab})-f(k\vec{r}_{ab})\sin(kx_{ab})].
\end{equation}

Besides the magnitude of $\Gamma$, the value of $\delta_c$ depends on two factors: 
The first factor $A(\phi,\tau)$ is a coefficient determined by the specific Rabi spectroscopy parameters i.e., $\Omega_0$, $t$ and $\delta$.
The second factor $F_{dd}$ is due to the finite electronic dipole coupling between the two clock states; since both $f(k\vec{r}_{ab})$ and $g(k\vec{r}_{ab})$ are present, the decay of the excited clock state and the interatomic dipole-dipole interaction both contribute of first order of $\Gamma$. 
In the Ramsey spectroscopy, of order $\Gamma$, the clock shift has the same structure as Eq.~(\ref{dc}); the first factor is different while the second one is still $F_{dd}$ \cite{ramsey}.

\begin{figure}[H]
\centering
\includegraphics[scale=0.6]{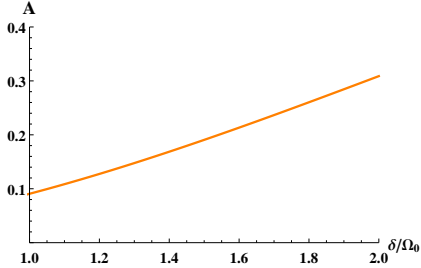}
\caption{The clock shift coefficient $A$ versus the reference detuing $\delta$ after a $\pi$-pulse.}\label{shiftcoefficient}
\end{figure}

To be specific, let us take a $\pi$-pulse, i.e., $2\Omega_0 t=\pi$, and choose $\delta/\Omega_0\approx 1.6$ where the excitation fraction in the absence of interactions equals $1/2$. With these parameters, the value of the coefficient is $A \approx 0.212712$. In the practically useful region where $|\delta/\Omega_0|$ is about between $1$ and $2$, Figure (\ref{shiftcoefficient}) shows that the coefficient $A$ varies around $0.2$.

The other factor $F_{dd}$ depends on the configuration of the lattice populated by the atoms. In the newly realized Fermi-degenerate three-dimensional optical lattice clock \cite{3dlatticejun}, three pairs of linearly polarised counter-propagating laser beams in each directions are used to produce the optical lattice, which is a primitive cubic lattice. The wave-vector magnitude of the lattice laser $k_L$ is roughly $k/1.07$ \cite{proceedings}.
For $N=10^5$ atoms unity filling in this cubic lattice, our numerical calculation yields $F_{dd}\approx 1.74682$.  Overall, $\delta_c/\Gamma\approx0.37157$ at the reference detuning $\delta/\Omega_0\approx 1.6$.
Thus for $^{87}$Sr, of which the clock transition wavelength is about $698$ nm and $\Gamma\approx 6\times 10^{-3}$ Hz, $\delta_c$ is at the relative level of $10^{-18}$. 

\begin{figure}[H]
\centering
\includegraphics[scale=0.18]{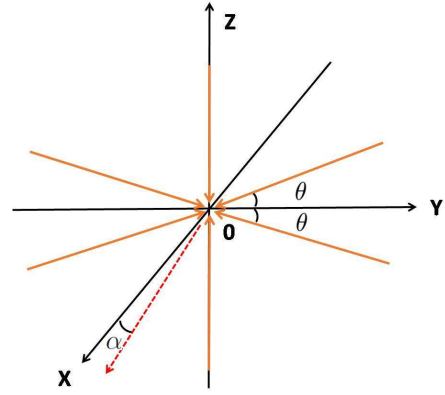}
\caption{Experimental setup for three dimensional optical lattice with six counter propagating laser beams. The four laser beams in xy plane has angle $\theta$ with Y axis. The dashed line represents the tuning of the driving laser orientation away from the positive $x$ direction with an angle $\alpha$. }\label{laser}
\end{figure}

It has been shown that by tuning the lattice configuration, one can reduce the value of $F_{dd}$ \cite{ramsey,optimizedgeometry}. For example as shown in Fig. \ref{laser}, based on the primitive cubic lattice configuration, one can vary the angle $2\theta$ between the two pairs of the counter-propagating laser beams in the $xy$ plane. The resulting lattice is a base-centered orthorhombic lattice (except for $\theta=\pi/4$ which is a tetragonal lattice). Another possible tuning is the relative angle between the driving laser and the lattice lasers. We parameterise the orientation tuning of the driving laser in the $xy$ plane in terms of $\alpha$, the angle between its orientation and the positive $x$ direction, which is represented by the dashed line in Fig.~\ref{laser}. 

For simplicity, let us first evaluate $F_{dd}$ as a function $\theta$ for $\alpha=0$. It is convenient to assume the following approximation for a large lattice sample \cite{ramsey}
\begin{equation}
\begin{aligned}
F_{dd}&=\frac{2}{N}\sum_{pairs}[g(k\vec{r}_{ab})\cos(kx_{ab})-f(k\vec{r}_{ab})\sin(kx_{ab})] \\
&=\frac{1}{N}\sum_{\vec{R}\neq 0, R<2r_0}U(k\vec{R})N(\vec{R}) \\
&=\frac{1}{N}\sum_{\vec{R}\neq 0,R<2r_0}U(k\vec{R})\int d\vec{r}\rho(\vec{r})\rho(\vec{r}+\vec{R})V,\label{fdda}
\end{aligned}
\end{equation} 
where the lattice vector difference $\vec{R}$ takes the values $m\vec a_1+n\vec a_2+l \vec a_3$ with $m,n,l$ integers and the lattice vectors $\vec a_1=(\pi/2k_L\sin\theta) \hat{e}_x+(\pi/2k_L\cos\theta)\hat{e}_y$, $\vec a_2=(\pi/2k_L\sin\theta) \hat{e}_x-(\pi/2k_L\cos\theta)\hat{e}_y$ and $\vec a_3=(\pi/k_L) \hat{e}_z$. The number of pairs of atoms separated by $\vec R$ is denoted by $N(\vec{R})$, and $U(k\vec{R})=g(k\vec R)\cos(k R_x)$. Note that 
since $f(k {\vec r}_{ab})\sin(kx_{ab})$ is an odd function of ${\vec r}_{ab}$, its summation drops out of Eq.~(\ref{fdda}).

\begin{figure}[H]
\centering
\includegraphics[scale=0.5]{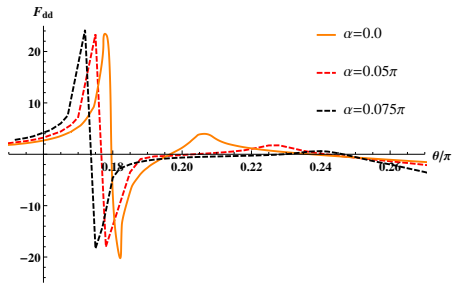}
\caption{$F_{dd}$ as a function of $\theta$ for various $\alpha$, the angle between the driving laser and the $x$ axis.}\label{fdd}
\end{figure}

For a spherical cloud of $N$ atoms populating the lattice with unity filling per site, the coarse-grained atomic density $\rho$ is uniform inside the cloud and equal to $1/V$ with $V$ the unit cell volume of the lattice, and is zero otherwise. The spherical cloud has a radius $r_0=\left(3NV/4\pi\right)^{1/3}$. The pair number $N(\vec{R})$ can be calculated with the density $\rho$ times the overlapping volume of two spheres whose centers are shifted with distance $R$. This overlapping volume equals
\begin{equation}
\begin{aligned}
&\int d\vec{r}\rho(\vec{r})\rho(\vec{r}+\vec{R})V^2\\
=&2\int_0^{\arccos(\frac{R}{2r_0})}\sin\psi d\psi\int_{\frac{R}{2\cos\psi}}^{r_0}dr r^2\int_0^{2\pi} d\phi\\
=&\frac{4\pi}3\left(r_0^3-\frac{3Rr_0^2}4+\frac{R^3}{16}\right).
\end{aligned} 
\end{equation}
Thus for $N=10^5$ atoms filling the lattice with unity, $F_{dd}$ as a function of $\theta$ is shown as the curve labeled $\alpha=0$ in Fig. \ref{fdd}.
  
Figure \ref{fdd} shows that $F_{dd}$ is rather sensitive to $\theta$ and can be suppressed or even cancelled for certain values of $\theta$. There are positions where the magnitude of $F_{dd}$ could very large which can be understood as due to the constructive Bragg scattering \cite{ramsey}. Meanwhile, there are also positions where $F_{dd}$ is close to zero as a result of the destructive Bragg scattering. These positions can be utilized in experiment to reduce the clock shift induced by the dipole-dipole interaction. We also examine cases of the small deviation angle $\alpha$ between the direction of driving laser and the x axis. Figure \ref{fdd} shows similar behaviors of $F_{dd}$ for small angle $\alpha\neq 0$ though the curves are shifted a bit in terms of $\theta$.  
 
\section{Summary} 
In summary, we study the clock shift in the Rabi spectroscopy due to the dipole-dipole interaction between atoms. We derive the analytic expression of the clock shift to first order of the interaction strength for three-dimensional optical lattice clocks. Numerically evaluation shows that the clock shift shall be at the $10^{-18}$ level for ${}^{87}$Sr though further suppression of the clock shift is possible by tuning the lattice configuration. Our result provides comparison with future experiments. 

\section*{Acknowledgements} 
We thank Donghong Xu for discussion. This work is supported by NSFC Grants No.
11474179, No. 11722438, and No. 91736103.

\end{document}